\documentclass[aps,prl,twocolumn,showpacs,superscriptaddress]{revtex4-2}
\usepackage{amsmath}
\usepackage{amssymb}
\usepackage{graphicx}
\usepackage{subfigure}
\usepackage{color}
\usepackage{float}
\usepackage{graphics}
\usepackage{hyperref}
\usepackage{txfonts}
\usepackage{tabularx}
\usepackage{array}
\usepackage{CJKutf8}

\newcommand{\CN}[1]{(\begin{CJK*}{UTF8}{gbsn}#1\end{CJK*})}
\begin{document}
\title{Higher-order Topological States in Chiral Split Magnons of Honeycomb Altermagnets}

\author{Xuan Guo\CN{郭璇}}
\altaffiliation{These two authors contributed equally to this work.}
\affiliation{Guangdong Provincial Key Laboratory of Magnetoelectric Physics and Devices, State Key Laboratory of Optoelectronic Materials and Technologies, Center for Neutron Science and Technology, School of Physics, Sun Yat-Sen University, Guangzhou, 510275, China}

\author{Meng-Han Zhang\CN{张梦菡}}
\altaffiliation{These two authors contributed equally to this work.}
\affiliation{Guangdong Provincial Key Laboratory of Magnetoelectric Physics and Devices, State Key Laboratory of Optoelectronic Materials and Technologies, Center for Neutron Science and Technology, School of Physics, Sun Yat-Sen University, Guangzhou, 510275, China}

\author{Dao-Xin Yao\CN{姚道新}}
\email{yaodaox@mail.sysu.edu.cn}
\affiliation{Guangdong Provincial Key Laboratory of Magnetoelectric Physics and Devices, State Key Laboratory of Optoelectronic Materials and Technologies, Center for Neutron Science and Technology, School of Physics, Sun Yat-Sen University, Guangzhou, 510275, China}

\begin{abstract}
We theoretically explore higher-order topological magnons in collinear altermagnets, encompassing a dimensional hierarchy ranging from localized corner modes to propagating hinge excitations. By employing antiferromagnetic interlayer coupling in bosonic Bogoliubov-de Gennes Hamiltonian, our work reveals anisotropic surface states and spatially distributed hinge modes propagating along facet intersections. We track the adiabatic evolution of Wannier centers to identify the bulk-polarization with second-order topological magnon insulator, where various magnon spectra demonstrate symmetry-protected band structure beyond conventional topology. Leveraging the stability and propagative properties of hinge modes, these unconventional magnons demonstrate manipulability in atomic-scale modifications of termination. Our study integrate altermagnetism with higher-order topology, which advance magnon-based quantum computing processing energy-efficient integrated architectures and information transfer.

\end{abstract}
\date{\today}
\maketitle

\begin{figure}[t]
\centering
{
\includegraphics[width=3.5 in]{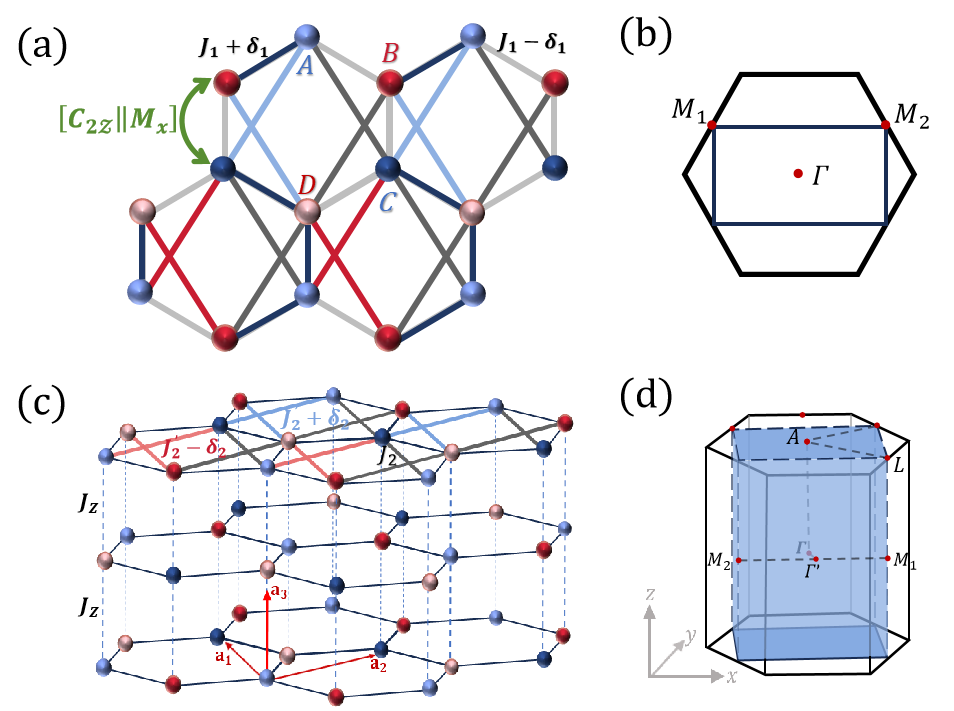}
}
\caption{(a) Schematic representation of a honeycomb altermagnet in the unit cell of antiferromagnetic interlayer coupling, where $A, B, C, D$ are four different sublattices within magnetic Ne$\acute{e}$l order. (b) and (d), Brillouin zones (BZ) for the two-dimensional lattice and three-dimensional structure. (c) The black bonds are the NNN ferromagnetic coupling with strength $J_{2}$, while the red(blue) bonds indicate Heisenberg interaction with strength $J_{2}-\delta_{2}$($J_{2}+\delta_{2}$). $a_{1}$, $a_{2}$ and $a_{3}$ are lattice vectors along with the sublattice sites.}
\label{Fig1}
\end{figure}

Collinear altermagnets are distinguished from conventional ferromagnets or antiferromagnets due to their distinct spin splitting patterns, which are symmetry-enforced in the absence of spin-orbit coupling \cite{altermagnet1,altermagnet2,altermagnet3,altermagnet4,altermagnet5,altermagnet7,altermagnet8}. The honeycomb lattice in altermagnetic systems reveals unconventional topological band structures\cite{altermagnethoneycombntopoband1,altermagnethoneycombntopoband2} and maintains magnetic two-fold rotational symmetry with antiparallel spin alignment.  We present higher-order topological magnons with bond-localized boundary modes, where unique alignment and symmetry properties influence the magnon spectra, leading to the emergence of exotic topological insulators and semimetals\cite{Higherorder1,HigherOrder3,qiu2025pairing,HigherOrder15,HigherOrder16,HigherOrder19}. Crucially, the different properties between bulk and these boundary modes in finite-sized systems arise from the onsite energy that is determined by the number of adjacent bonds. The nearest-neighbor (NN) adjustment parameter $\delta_{1}$ and next-nearest-neighbor (NNN) adjustment parameter $\delta_{2}$  break the inversion symmetry, lifting Kramers degeneracy to distinguish magnon modes with opposite chirality\cite{chiralspiltmagon,BrokenKramersdegeneracy1,BrokenKramersdegeneracy2}. Hence the bulk spectrum exhibits a finite energy gap separating the acoustic and optical branches, in contrast to the gapless Dirac point characteristic of conventional graphene. Dimensional anomalies indicate second-order topological magnon insulator (SOTMI) extending beyond conventional $\mathbb{Z}_2$ topology\cite{z2-1,z2-2}. Experimentally, honeycomb lattice altermagnets can be probed at the single-atom scale using scanning tunneling microscopy (STM), which can directly observe the local spectral signals of corner modes and the transport properties of hinge states\cite{stm2}. 

We propose the emergence of intrinsic corner states remaining robust against small perturbations and various edge terminations, which propel the realization of magnonic quantum devices capable of supporting robust information transmission channels\cite{Corner3,Corner8,magnonicquantumdevices}. The honeycomb lattice in altermagnetic systems reveals unconventional topological band structures, introducing a symmetry operation that combines effective time-reversal symmetry with additional lattice symmetry of the magnetic order. Based on antiferromagnetic (AFM) interlayer coupling, stable higher-order topological magnon states can be constructed, which combine controllability with interference-resistance. Experimental detection of localized magnon corner modes via near-field Brillouin light scattering \cite{Brillouinscattering} is feasible, where the higher-order skin effect manifests itself through the existence of magnon corner modes in extended dimensions, where these modes remain robust against disorder. The asymmetric Brillouin-zone (BZ) distribution of magnon states permits novel transport phenomena through higher-dimensional topological boundaries, enabling topological circuit designs combining both low-energy processing and robust data transmission\cite{circuitry1,circuitry2}. 

Leveraging the intrinsic altermagnetic order of the magnon Hamiltonian, we demonstrate an effective topological phase transition characterized by robust corner-localized modes, marking a shift from a conventional topological magnon phase to a higher-order topological phase\cite{phase}. This transition surpasses the conventional bulk-boundary correspondence, manifesting a nodal phase under AFM interlayer coupling, despite the absence of non-collinear ordering or interaction asymmetry. The bosonic Bogoliubov-de Gennes (BdG) Hamiltonian captures low-dissipation magnetic excitations as propagating hinge modes with distinct chiral split magnon bands, where Wannier spectra reveal bulk polarization along the stacking direction. We unveil these hinge modes traversing a defect-free trajectory through the junctions where facets meet, because the spectral gap energetically isolates the topologically protected state from decoherence channels. Our work provides a theoretical foundation for magnonic circuits with resilience to disorder, supporting signal integrity during information processing and tunability in atomic-scale engineering.


\begin{figure*}[t]
\centering
{
\includegraphics[width=7.2 in]{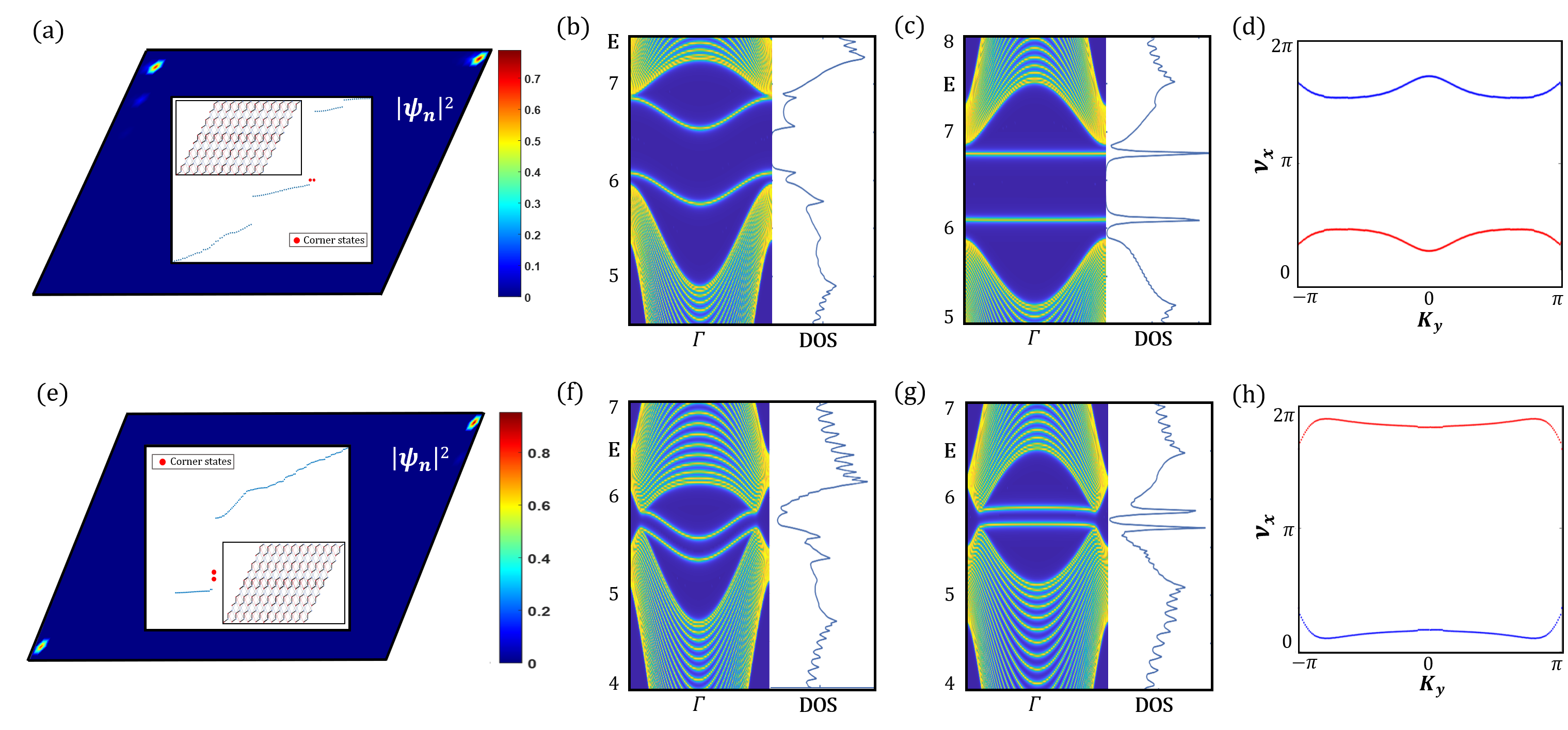}}
\caption{ (a) and (e) depict the magnon eigenvector probabilities $|\psi_{n}|^2$ of corner modes for $J_{2}'/J_{2}$=0.8, while $\delta_{1}$ and $\delta_{2}$ are both -0.4. (b) and (c) are edge spectra and local density of states (DOSs) for the zigzag and armchair nanoribbon respectively, where the parameters are chosen as before. (d) The Wannier center evolution is depicted as a function of $k_{y}$. Panels (f) and (g) illustrate the gapped altermagnetic honeycomb model with $J_{2}'/J_{2}$=1.2, $\delta_{2}$=-0.4 and $\delta_{1}$=-0.1, identifying the intrinsic second-order topological phase. 
(h) The Wannier center evolution corresponding to the nodal-line semimetal.}
\label{Fig2}
\end{figure*}

\textit{Model and method.} In $N\acute{e}el$-ordered spin-split antiferromagnets, we calculate the magnetic excitations featuring tetra-spin unit cells as shown in Fig.~\ref{Fig1}. The magnonic system is described by a Heisenberg Hamiltonian incorporating four distinct sublattices, adopting the bosonic BdG formalism.

\begin{align}
\begin{split}
H &= (J_{1} \pm \delta_{1}) \sum_{\langle i,j \rangle_{1}} \mathbf{S}_i \cdot \mathbf{S}_j +J_{z} \sum_{\langle i,j \rangle_{2}} \mathbf{S}_i \cdot \mathbf{S}_j\\
&+ \sum_{{\langle\langle i,j \rangle\rangle}} \left[J_2 \mathbf{S}_i \cdot \mathbf{S}_j + (J_2' \pm \delta_{2}) \mathbf{S}_i \cdot \mathbf{S}_j\right],\\
\end{split}\label{Eq1}
\end{align} 

where $J_{1}\pm\delta_{1}$ denote the nearest neighbor couplings, while $J_2$ and $J_{2}'\pm\delta_{2}$ correspond to distinct variants of next nearest neighbor couplings. The spin-magnon transformation is implemented through Holstein-Primakoff representation followed by Fourier transformation. The magnon bands can be diagonalized via a Bogoliubov transformation from its eigenvectors $\psi^{}_{\boldsymbol{k}}$=$T^{\dag}\phi^{}_{\boldsymbol{k}}$, while the para-unitary transformation $T$ provides a robust mechanism to study chiral split magnons. Altermagnetism can be described within a generalized BdG equation formalism, employing Pauli matrices $\sigma$ and $\tau_{z}$ that act on the lattice degrees of freedom, respectively. We have $T(\sigma_{0}\otimes\tau_{z})T^{\dag}=\sigma_{0}\otimes\tau_{z}$, where $\sigma_{0}\otimes\tau_{z}$ operator guarantees the magnon BdG Hamiltonian maintaining particle-hole symmetry. The left (right)-hand chirality of magnon bands is denoted as $\langle0|\psi_{k}\mathbf{S}^{z}\psi_{k}^{\dag}|0\rangle$=$\pm 1$, satisfying the generalized orthonormal condition $\langle\langle\psi_{m}|\psi_{n}\rangle\rangle$=$\psi_{m}^{\dag}(\sigma_{0}\otimes\tau_{z})$ $\psi_{n}$. We introduce the interlayer coupling $J_{z}$ in Fig.~\ref{Fig1}(c), where distinct topological phases emerge as the ratio $J_{2}'/J_{2}$ is varied.

The emergence of chiral split Weyl magnonic states stems from strategic manipulation of bulk-embedded nodal lines, which is enabled by the non-zero $\delta_{2}$ in the $J_{2}'/J_{2}<1$ parameter regime. When arranged in the stacking configuration, pairwise Weyl points exhibit evolution along high-symmetry paths ($\Gamma-L$). Considering the composite symmetry of altermagnetism, the Berry connection demonstrates a modified expression via the BdG inner product of the transformed quasi-particle wavefunctions, given by $\mathbf{A}_{n}(\mathbf{k})=i[(\sigma_{0}\otimes\tau_{z})T^{\dag}(\sigma_{0}\otimes \tau_{z})\nabla T]_{nn}$. The nontrivial topological characteristics are quantified by a $\pi$ Berry phase, originating from the contour integration of Berry connection along closed loops surrounding Weyl nodes in momentum space. By expanding the unit cell, we derive a reduced BZ and systematically analyze the wavefunction parity at time-reversal invariant momenta (TRIM) points $\Gamma_{i} = (\Gamma, M)$ using the inversion operator $\mathbf{P} = \sigma_{0}\tau_{x}$. When the exchange coupling ratio $J_{2}'/J_{2}$ exceeds unity, the parity eigenvalues unequivocally demonstrate the topological robustness of chiral split nodal lines\cite{altermagnethoneycombntopoband1}, solidifying their protected nature against symmetry-preserving perturbations.  

\begin{equation}
\begin{split}
(-1)^{\nu}=\prod_{n}\prod_{i}\xi_{n}(\Gamma_{i}).\\
\end{split}
\end{equation}

Under finite $k_z$, we enable layer-resolved $\mathbb{Z}_2$ classification via topological invariant $\nu$ for trivial or topological states with the presence of 2-fold spin rotation symmetry $C_{2z}$. In the insets of Fig.~\ref{Fig1}(a), the corner-localized modes are inherently stabilized by magnetic mirror symmetry, protected by their $\mathbb{Z}_2$ topological nature quantified by the $\nu$ index. The effective time-reversal symmetry operator $\mathcal{T} = \sigma_{z} \otimes i\tau_{y}\mathcal{K}$ enables nodal lines and Weyl magnons in the magnonic band structure with $\mathcal{K}$ representing the complex conjugation operation. We examine the symmetry properties of wavefunctions associated with the parity eigenvalues of $\mathbf{P}$ are $\xi_{n}(\Gamma_{i})$ at $\Gamma_{i}$. Our theoretical investigation, employing a four-sublattice bond dimerization approach, uncovers the $\mathbb{Z}_2$ topology analogous to a two-dimensional Su-Schrieffer-Heeger (SSH) conﬁguration\cite{dimerization1,ssh1,ssh2}, which features an antiferromagnetic order with staggered interlayer coupling. As shown in Fig.~\ref{Fig2}, the altermagnetic honeycomb lattice undergoes a transition from $\mathbb{Z}_2$ classification to higher-order topology because of the emergence of a substantial energy gap in the Kramers degeneracy\cite{z21,z22,z23}.

\textit{Corner Modes in 2D. } Preserving the combined symmetry with magnetic two-fold rotation $\mathcal{C}_{2}$ and $\mathcal{T}$, the inherent bipartite nature of honeycomb lattice offers altermagnetism for studying topological magnons\cite{bipartite}. The $\nu$ = 1 for the hexagonal lattice monolayer indicates a 2D $\mathbb{Z}_2$ topological phase, in which a significant gap induced by $\delta_{2}$ emerges in the originally degenerate chiral split nodal lines. We study the SOTMI hosting localized corner states by implementing open boundary conditions and numerical diagonalization with $J_z=0$. The magnon corner states emerge due to the nearest-neighbor bond alternations of 2D lattice Hamiltonian, possessing non-zero $\delta_1$ value with well-defined in-gap states, whose spectral signatures are shown in Fig.~\ref{Fig2}(b, c). We consider the corners in the presence of $\delta_{1}$, where the effective time-reversal symmetry is broken and the edge modes are gapped. When two gapped zigzag boundaries meet at a corner, they induce localized corner states characterized by two pairs of degenerate energy levels. The on-site potential causes the gap in edge state generated by $\delta_{1}$ to shift into the gap in bulk energy, ultimately creating in-gap states within the band structure of zigzag-terminated nanoribbons as shown in Fig.~\ref{Fig2}(b). Additionally, the magnonic density of states (DOSs) manifest sharp van Hove singularities as a spectroscopic signature for detecting topological properties within the magnon band structure. Fig.~\ref{Fig2}(c, g) presents the calculated magnon energy spectrum as a function of eigenvalue index. Notably, the two in-gap magnon corner states persist with negligible energy shift, demonstrating their remarkable robustness against local perturbations. This stability is further confirmed by introducing a point defect at the top corner of the rhombus-boundary sample, where the corner states remain spectrally isolated within the bulk gap\cite{defect}.

As shown in Fig.~\ref{Fig2}(a, e), we numerically calculate and visualize both the magnon dispersion spectrum and $|\psi_{n}|^2$ corresponding to the two emergent in-gap states. These topological states exhibit remarkable stability, maintaining strong spatial localization near the original corner sites throughout the simulation. Coinciding with the normalized site-resolved distribution of bulk polarization\cite{bulkpolarization}, they exhibit proximity to the lower bulk states while gradually approaching the upper bulk states as the coupling ratio varies under the condition $\delta_1 < 0$. Distinct from other gap states that lack zero-dimensional localization, the corner states highlight the topological and altermagnetic origins of higher-order magnon insulator. We numerically track the evolution of Wannier centers by the Wilson loop operator, which is constructed from the $n$th Bloch eigenstates $|\psi_{n, k_{x}, k_{y}}\rangle$ in the occupied subspace. The Wilson loop operator is constructed through the ordered product of overlap matrix between adjacent Bloch wavefunctions $\langle\langle\psi_{n, k_{x}, k_{y}+\Delta k_{y}}|\psi_{n, k_{x}, k_{y}}\rangle\rangle$, where $\Delta k_{y}$ quantifies the discretization. Bulk polarization is determined by averaging the Wannier center during the BZ sampling, which subsequently generates robust corner modes whose spatial localization demonstrates notable stability against various perturbations. The symmetric distribution of Wannier spectrum indicates the absence of the first-order topological phase as illustrated in Fig.~\ref{Fig2}(d, h), since topological edge states typically break such mirror symmetry patterns. Despite thermal perturbations introducing noise, the overall topological invariants remain robust against local perturbations ensuring corner-state persistence in SOTMI\cite{topologicalinsulator3,topologicalinsulator9,Corner7}. Quantitative analysis reveals their probability density profiles preserve exact mirror symmetry about the central axis, with characteristic exponential decay away from the boundary corners. The persistent spatial localization and protected symmetry properties suggest these in-gap states may serve as robust quantum information carriers in magnonic systems.

\begin{figure}[t]
\centering
{

\includegraphics[width=3.5 in]{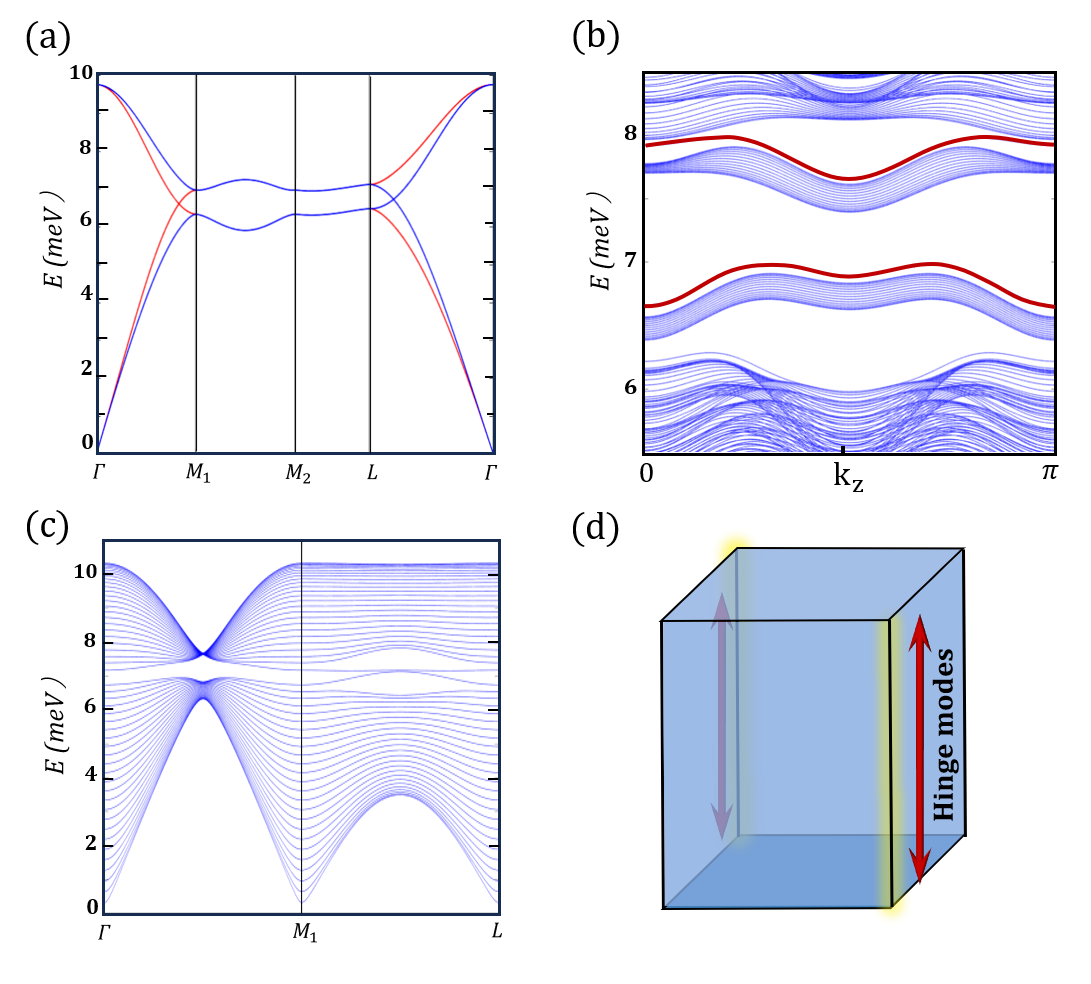}}

\caption{(a) The bulk magnon band structure for $J_{z}=0.8J_{1}$, where the red (blue) bonds indicate right (left)-hand chirality. (b) Numerical simulations are performed on a finite-size lattice with dimensions $N_{x} \times N_{y} =20 \times 20$, revealing localized hinge modes that propagate unidirectionally along the edges. These topologically protected hinge modes are explicitly highlighted in red within the spectral plot, emphasizing their distinct energy dispersion relative to the bulk bands. (c) The magnon surface states in the $k_{x}-k_{z}$ plane with $N_{y}$ = 20 along y direction. (d) The hinge modes are marked by red arrows along the sample edges, indicative of topologically protected states. }
\label{Fig3}
\end{figure}

\textit{ Surface States in 3D. } Altermagnetic honeycomb lattices exhibit weak topological phases through intrinsic mirror symmetry $\mathbf{M}=\sigma_{x}\tau_{x}$, stabilizing hinge modes with anisotropic gapless topological surface states with the AFM interlayer coupling as shown in Fig.~\ref{Fig3}(d). Characterized by two distinct eigenvalues ($\pm1$) of $\mathbf{M}$, the sector decomposition in distinct modes induces band degeneracy via altermagnetism in a three-dimensional system as illustrated in Fig.~\ref{Fig3}(a). Since mirror symmetry remains independent of $k_{z}$, the interlayer coupling terms enforce identical sectoral components  to remain consistent across the decomposed sectors. We delve into the surface states accounting for finite layer thickness along the stacking axis depicted in Fig.~\ref{Fig3}. The three-dimensional architecture reveals an intrinsic asymmetry between side surfaces as their zigzag edge terminations belong to different sublattice species, with an inversion symmetry present in $\delta_{2}$.The surface Dirac cone emerges in the gap along the $k_{z}$ axis, where gapless topological surface states would exist in the projected $k_{x}-k_{z}$ plane with ﬁnite width along the y axis, diﬀerent from the zigzag-terminated surfaces that are gapped. When projected onto the $k_{x}-k_{z}$ plane, the spatial inversion creates gapless surface states emerging parallel to the stacking direction, where nontrivial surface states emerge on one facet and the opposing facet retains trivial topology as shown in Fig.~\ref{Fig3}(c). A weak topological phase aligned with the stacking axis with $[\nu_0, (\nu_1, \nu_2, \nu_3)]=[0, (0, 1, 0)]$, which is robust and exhibits non-trivial topological properties upon localization on surfaces perpendicular to the stacking axis. The introduction of $\delta_{1}$ and $\delta_{2}$ breaks the $\mathcal{T}$ while maintaining spatial symmetries such as mirror reflection, consequently inducing a higher-order topological insulating phase characterized by robust hinge modes\cite{Higher-orderphase,Topologicalinsulators1,j1}. 

In the AA-type stacking geometry, we systematically visualize the magnonic band structure projected along the $k_{x}$-$k_{y}$ plane while maintaining precise control over the spectral weight of $k_{z}$ component. A particularly intriguing scenario arises when examining AFM interlayer Heisenberg coupling between two zigzag-terminated surfaces interconnected via an armchair-type junction. In the insets of Fig.~\ref{Fig3}(b), this unique configuration gives rise to the emergence of distinct in-gap modes within the surface bandgap region, suggesting novel topological magnonic states at the interface.  The energy gap at the intersection of two surface states generates localized hinge modes with tunable magnetic properties. During a gap-closing transition, a second trivial bulk gap emerges between the acoustic and optical bands as the sign of $\delta_{1}$ reverses. To distinguish the higher-order topological phase from the trivial gapped state, we systematically evaluate the bulk polarization of the magnon Hamiltonian across all occupied bands. The second-order topological phase characterized by robust hinge modes persists uniformly at arbitrary $k_{z}$ momenta, even in the presence of symmetry-breaking disorder. Such hinge modes can be interpreted as a generalized variant of the fundamental SSH model, exhibiting non-chiral propagation on surfaces perpendicular to the stacking axis, as shown in Fig.~\ref{Fig3}(d). The altermagnetic lattice geometry introduces additional degrees of freedom with a characteristic dimerization pattern, while the stacking symmetry creates a rich platform for studying engineered magnon dispersion and transport properties beyond the conventional topology in artificially structured quantum materials. 


\textit{Conclusion.} We study the interplay between altermagnetic ordering and the dimensional hierarchy of interlayer AFM coupling in honeycomb lattice, demonstrating a novel mechanism for generating higher-order topological magnon excitations. The magnon corner states in 2D honeycomb altermagnets are governed by the coordination number of adjacent magnetic bonds, which induces a stark distinction between the bulk magnon bands and topologically protected boundary modes in finite-size systems. The interplay of anisotropic exchange interactions and lattice geometry breaks conventional symmetry constraints, where the nested Wilson loops quantify the nontrivial bulk polarization. Additionally, AA stacked honeycomb lattices offer a promising avenue for engineering exotic 1D hinge modes, enabling a magnonic analog of electronic higher-order topology with hierarchy of dimensional reduction. The magnon dispersion spectra remain twofold degeneracy along the high-symmetry line at $k_x = 0$, which indicates unique spin-split band structure of altermagnetism relevant to centrosymmetric systems. For AA-type stacking, the preserved mirror symmetry enforces protection mechanism in generating intrinsic higher-order magnons beyond conventional band topology. Combining crystalline symmetry-protected topology with dimensionally confined spin-wave propagation can lift the magnon Kramers degeneracy and generate hinge modes. This topologically protected hinge states can enhance the stability of signal transmission and have inherent resistance to interference factors such as disorder and impurities, thereby advancing the precision-engineered design of low-dissipation magnon devices.

We theoretically confirm a novel altermagnetic lattice with two and three-dimensional integrated architectures exceeding the limits of conventional bulk-boundary correspondence. Our proposed architecture maintains signal integrity, while the termination-dependent modulation provides a versatile platform for reconfigurable magnonic device implementations. Thus, we proceed to higher-order topological magnons with topological features arising from the chirality of band structure, which enable robust corner and hinge states. Monolayer antiferromagnets possessing space-time inversion symmetry can be effectively transformed into an altermagnetic state through external stimuli such as applied electric fields\cite{MonoMnPX3}. Two prototypical van der Waals materials can construct three-dimensional stacked honeycomb lattices within the weak interlayer coupling $J_{z}$ by local moments of Mn, which can be experimentally observed via inelastic neutron scattering measurements\cite{MnPX3}. Distinct magnetic exchange coupling parameters are obtained as $J_{1}$=-0.65 meV, $J_{2}$=-0.03 meV, and $J_{z}$ = -0.0019 meV for $MnPS_{3}$, while the corresponding values are $J_{1}$=-0.40 meV, $J_{2}$=-0.03 meV, and $J_{z}$ = -0.031 meV for $MnPSe_{3}$\cite{MnPS3,MnPSe3}. For practical application, the higher-order skin effect, which exhibits a stronger degree of boundary localization and robustness, enables the development of quantum sensors capable of highly sensitive detection of extremely weak signals\cite{ultrasensitive}. Furthermore, the artificial spin lattice technique provides an atomic-precision experimental platform for bosonic systems, where ultra-low-temperature vector magnetic field can induce higher-order topological phase transitions\cite{esrstm}.

\textit{Acknowledgements.}
We thank Lu Xiao for helpful discussions. This work was supported by the National Key Research and Development Program of China (Grant No. 2022YFA1402802), the National Natural Science Foundation of China (Grant Nos. 92165204 and 12494591), Guangdong Provincial Key Laboratory of Magnetoelectric Physics and Devices (Grant No. 2022B1212010008), Research Center for Magnetoelectric Physics of Guangdong Province (Grant No. 2024B0303390001), and Guangdong Provincial Quantum Science Strategic Initiative (Grant No. GDZX2401010).


\bibliography{main}

\end{document}